# LU transformation invariant operators and LU transformation invariant

Xin-wei ZHA [1] and Chun-min ZHANG [2]

1. Department of Applied Mathematics and Physics, Xi'an Institute of Posts and Telecommunications, Xi'an, 710061 Shaanxi, China
2. Shool of Science, Xi'an Jiaotong University, Xi'an, 710049 Shaanxi, China

**Abstract:**

We proposed a concept of LU transformation invariant operators. By using this operator, arbitrary multi-qubit states LU transformation invariant and SLOCC invariant could be easily obtained. And we find that presences two kinds of invariant operators and corresponding invariants. One kind of operators yields LU invariants and the other operators results in SLOCC invariants. For three-qubit states, all independence LU transformation invariant are obtained. Furthermore, by this system method, arbitrary multi-qubit states invariants can be given.

## 1. Introduction

The determination of a given state is entangled is a fundamental and very important topic in the field of the quantum information theory, therefore, it is obvious that the invariants of states, under unitary transformations which act on single particles separately ("local" transformations), playa an essential role[1~4] because it could present the finest discrimination between different types of entanglement. Which can be regarded as coordinates on the space of entanglement types (equivalently, the space of orbits of the group of local transformations). The state of art of the current studying, many researchers have conducted the examination on such invariants by different methods, in terms of density matrices [1] and multidimensional determinants[5]. However, there is not a systematic method to look for LU transformation invariant, especially for SLOCC invariant.

The alternative approach to the investigation for pure multi-qubit states, we have studied on some of the recently work[6~9] on entanglement and introduce a concept of LU transformation invariant operators, by which arbitrary multi-qubit LU transformation invariants and SLOCC invariant can be easily obtained.

## 2. LU transformation Invariants operator of multi-qubit states

We know that two n-party pure states $|\psi\rangle, |\varphi\rangle$ are equivalent under LU transformations if

$$|\psi\rangle = U_1 \otimes U_2 \otimes \cdots \otimes U_n |\varphi\rangle \qquad (1)$$

where $U_j$ is an unitary operation acting on the Hilbert space of the jth party:

$$U = \begin{bmatrix} e^{i\alpha} & 0 \\ 0 & e^{-i\alpha} \end{bmatrix} \begin{bmatrix} \cos\omega & \sin\omega \\ -\sin\omega & \cos\omega \end{bmatrix} \begin{bmatrix} e^{i\beta} & 0 \\ 0 & e^{-i\beta} \end{bmatrix}$$

or, $$U = e^{i\alpha\sigma_z} e^{i\omega\sigma_y} e^{i\beta\sigma_z} \qquad (2)$$

**Definition 1** For arbitrary Pauli matrix operator we define a LU transformation operator

$$\hat{\sigma}_i' = U^{-1}\hat{\sigma}_i U \qquad i = 0, x, y, z \qquad (3)$$

Under this definition for an arbitrary function $f$, if

$$f(\hat{\sigma}_i') = f(\hat{\sigma}) \qquad (4)$$

we call $f(\hat{\sigma})$ LU transformation `first` invariant operator.

Obviously, the identity operator $\hat{\sigma}_0 = \hat{I}$ is a LU transformation invariant operator. The Pauli matrix operators and $\hat{\sigma}_i$ ($i = x, y, z$) in the contrary are not individually LU transformation invariant operators, but we have

$$\hat{\sigma}_i' = \sum_{j=x,y,z} O_{ij} \hat{\sigma}_j \qquad (5)$$

where $O_{ij}$ is the corresponding real orthogonal matrix in SO(3). Therefore $\hat{\sigma}_x^2 + \hat{\sigma}_y^2 + \hat{\sigma}_z^2$ is a LU invariant operator. More generally, we can construct other LU transformation invariant operators

**Definition 2**

let $$\hat{\sigma}_i'' = U^{-1}\hat{\sigma}_i U^* \qquad (6)$$

If $f(\hat{\sigma}'') = f(\hat{\sigma})$, We call $f(\hat{\sigma})$ second LU transformation invariant operator.

Obviously, $\hat{\sigma}_y$ is a second LU transformation invariant operator.

On the other hand, we obtain

$$\hat{\sigma}_x''^2 + \hat{\sigma}_z''^2 - I''^2 = \hat{\sigma}_x^2 + \hat{\sigma}_z^2 - I^2 \tag{7}$$

So $\hat{\sigma}_x^2 + \hat{\sigma}_z^2 - I^2$ is a second LU transformation invariant operator. More generally, we can construct other second LU transformation invariant operators

## 3. Invariants of multi-qubit states

By two kinds of LU transformation invariant operator, we can easily obtain two kinds of invariants. By first invariant operator, we can obtain real number invariant(LU transformation invariant); By second invariant operator, we can obtain complex number invariant(SLOCC transformation invariant).

Now we will examine some invariants of an arbitrary multi-qubit state with the LU transformation invariants operators.

### 3.1 The real number LU transformation invariant

In fact, the invariant:

$$I_i = 4\det\rho_i = 2(1 - tr\rho_i^2) \tag{8}$$

can be obtained by the LU transformation invariant operators $\hat{\sigma}_0 = \hat{I}$ and $\hat{\sigma}_x^2 + \hat{\sigma}_y^2 + \hat{\sigma}_z^2$

$$I_i = \langle\psi|\psi\rangle - \langle\psi|\sigma_{ix}|\psi\rangle^2 - \langle\psi|\sigma_{iy}|\psi\rangle^2 - \langle\psi|\sigma_{iz}|\psi\rangle^2 \tag{9}$$

where $i = 1, 2\cdots$ and $\rho_i$ are density matrices.

Similarly, we can show that

$$\begin{aligned}I_{ij} = \langle\psi|\psi\rangle - [&\langle\psi|\hat{\sigma}_{ix}\hat{\sigma}_{jx}|\psi\rangle^2 + \langle\psi|\hat{\sigma}_{ix}\hat{\sigma}_{jy}|\psi\rangle^2 + \langle\psi|\hat{\sigma}_{ix}\hat{\sigma}_{jz}|\psi\rangle^2 \\ +&\langle\psi|\hat{\sigma}_{iy}\hat{\sigma}_{jx}|\psi\rangle^2 + \langle\psi|\hat{\sigma}_{iy}\hat{\sigma}_{jy}|\psi\rangle^2 + \langle\psi|\hat{\sigma}_{iy}\hat{\sigma}_{jz}|\psi\rangle^2 \\ +&\langle\psi|\hat{\sigma}_{iz}\hat{\sigma}_{jx}|\psi\rangle^2 + \langle\psi|\hat{\sigma}_{iz}\hat{\sigma}_{jy}|\psi\rangle^2 + \langle\psi|\hat{\sigma}_{iz}\hat{\sigma}_{jz}|\psi\rangle^2]\end{aligned} \tag{10}$$

are also LU transformation invariants.

and
$$I_i + I_j + I_{ij} = 4(1 - tr\rho_{ij}^2) \tag{11}$$

Similarly, we can give other high degree LU transformation invariant.

### 3.2 The complex number SLOCC transformation invariant

For second LU transformation invariant of arbitrary multi-qubit, the construction is more complex than first. For arbitrary even multi-qubit, there are

$$C_{12\cdots n} = \langle \psi | T_1 T_2 \cdots T_n | \psi^* \rangle \qquad (12)$$

where $T = i\sigma_y$.

For arbitrary odd multi-qubit, there are

$$Z_{12\cdots n-1} = \langle \psi | T_1 T_2 \cdots T_{n-1} \sigma_{nx} | \psi^* \rangle^2 + \langle \psi | T_1 T_2 \cdots T_{n-1} \sigma_{nz} | \psi^* \rangle^2 - \langle \psi | T_1 T_2 \cdots T_{n-1} | \psi^* \rangle^2 \qquad (13)$$

Certainly, we can give other SLOCC transformation invariant.

### 4. Results and discussion

For a given pure state, and $\langle \psi | \psi \rangle = 1$; there are $2^{n+1} - (3n+1)$ invariant parameters, For pure three-qubit states, According to A. Sudbery[1] conclusion, there are six algebraically independent local invariants.

$$I_1 = \langle \psi | \psi \rangle$$

$$I_2 = tr(\rho_C^2)$$

$$I_3 = tr(\rho_B^2)$$

$$I_4 = tr(\rho_A^2)$$

$$I_5 = 3tr[(\rho_A \otimes \rho_B)\rho_{AB}] - tr(\rho_A^3) - tr(\rho_B^3)$$

$$I_6 = \tau_{ABC} = 4|d_1 - 2d_2 + 4d_3|$$

By two kinds of LU transformation invariant operator, we can easily obtain two kinds of LU transformation invariants.

$$I_1 = \langle \psi | \psi \rangle$$

$$I_2 = tr(\rho_C^2) = \frac{1}{2}\left(\langle \psi | \psi \rangle + \langle \psi | \sigma_{Cx} | \psi \rangle^2 + \langle \psi | \sigma_{Cy} | \psi \rangle^2 + \langle \psi | \sigma_{Cz} | \psi \rangle^2 \right)$$

$$I_3 = tr(\rho_B^2) = \frac{1}{2}\left(\langle \psi | \psi \rangle + \langle \psi | \sigma_{Bx} | \psi \rangle^2 + \langle \psi | \sigma_{By} | \psi \rangle^2 + \langle \psi | \sigma_{Bz} | \psi \rangle^2 \right)$$

$$I_4 = tr(\rho_A^2) = \frac{1}{2}\left(\langle\psi|\psi\rangle + \langle\psi|\sigma_{Ax}|\psi\rangle^2 + \langle\psi|\sigma_{Ay}|\psi\rangle^2 + \langle\psi|\sigma_{Az}|\psi\rangle^2\right)$$

$$I_5 = 3tr[(\rho_A \otimes \rho_B)\rho_{AB}] - tr(\rho_A^3) - tr(\rho_B^3)$$
$$= \frac{1}{4}(\langle\psi|\psi\rangle + 3\Delta_{AB})$$

where,

$$\Delta_{AB} = \langle\sigma_{AX}\rangle\langle\sigma_{BX}\rangle\langle\sigma_{AX}\sigma_{Bx}\rangle + \langle\sigma_{AX}\rangle\langle\sigma_{By}\rangle\langle\sigma_{Ax}\sigma_{By}\rangle + \langle\sigma_{AX}\rangle\langle\sigma_{Bz}\rangle\langle\sigma_{Ax}\sigma_{Bz}\rangle$$
$$+ \langle\sigma_{Ay}\rangle\langle\sigma_{Bx}\rangle\langle\sigma_{Ay}\sigma_{Bx}\rangle + \langle\sigma_{Ay}\rangle\langle\sigma_{By}\rangle\langle\sigma_{Ay}\sigma_{By}\rangle + \langle\sigma_{Ay}\rangle\langle\sigma_{Bz}\rangle\langle\sigma_{Ay}\sigma_{Bz}\rangle \quad (14)$$
$$+ \langle\sigma_{Az}\rangle\langle\sigma_{Bx}\rangle\langle\sigma_{Az}\sigma_{Bx}\rangle + \langle\sigma_{Az}\rangle\langle\sigma_{BY}\rangle\langle\sigma_{Az}\sigma_{By}\rangle + \langle\sigma_{Az}\rangle\langle\sigma_{Bz}\rangle\langle\sigma_{Az}\sigma_{Bz}\rangle$$

and $\langle\sigma_{Ax}\rangle = \langle\psi|\sigma_{Ax}|\psi\rangle$, etc.

$$I_6 = \tau_{ABC} = 4|d_1 - 2d_2 + 4d_3| = |C_{AB}|$$

where

$$C_{AB} = \langle\psi^*|\sigma_{1y}\sigma_{2y}\sigma_{3x}|\psi\rangle^2 + \langle\psi^*|\sigma_{1y}\sigma_{2y}\sigma_{3z}|\psi\rangle^2 - \langle\psi^*|\sigma_{1y}\sigma_{2y}|\psi\rangle^2 \quad (15)$$

Furthermore, we find $C_{AB} = C_{AC} = C_{BC}$.

the above idea we can obtain the LU transformation invariants for any multi-qubit pure and mixed states. Further discussion we will given in other papers.

**Acknowledgement**

The authors gratefully acknowledge the support of Chinese Natural Science Foundation. This work is supported by the Chinese Natural Science Foundation under Contract Nos. 40375010，60278019, and by the Science and Technology Plan Foundation of Shaanxi Province under Contract Nos. 2001K06-G12, 2004A15 and Science Plan Foundation of office the Education Department of Shaanxi Province Contract Nos. 05JK288, We thank Yong-De Zhang and Sixia Yu for useful discussion. We are thankful to Dr. mingjun zhao for his help with this paper.